\documentclass{jpp}
\usepackage{amsmath}
\usepackage{amssymb}
\usepackage{xcolor}
\usepackage{physics}
\usepackage{subcaption}
\usepackage{siunitx}								
\usepackage{overpic}								
\usepackage{booktabs}                               
\usepackage{multirow}
\usepackage{hyperref}

\usepackage{dirtytalk} 
\usepackage{mathtools} 
\usepackage{cancel} 

\newcommand{\figref}[1]{\mbox{fig.~\ref{#1}}}
\newcommand{\tabref}[1]{\mbox{table~\ref{#1}}}
\newcommand{\secref}[1]{\mbox{sec.~\ref{#1}}}

\newcommand{\appref}[1]{\mbox{appendix~\ref{#1}}} 
\renewcommand{\eqref}[1]{\mbox{eq.~(\ref{#1})}}

\newcommand{\twosecref}[2]{\mbox{secs.~\ref{#1}~and~\ref{#2}}}
\newcommand{\twofigref}[2]{\mbox{figs.~\ref{#1}~and~\ref{#2}}}

\newcommand{\figpanel}[2]{fig.~\hyperref[#1]{\ref*{#1}(#2)}}
\newcommand{\figpanels}[3]{fig.~\hyperref[#1]{\ref*{#1}(#2)--(#3)}}
\newcommand{\figpanelNoPrefix}[2]{\hyperref[#1]{\ref*{#1}(#2)}}

\newcommand{\divJ}[1]{\boldsymbol{\nabla} \cdot #1}
\newcommand{\curlJ}[1]{\boldsymbol{\nabla} \times #1}
\newcommand{\dirdvJ}[1]{#1 \cdot \boldsymbol{\nabla}}

\newcommand{\pd}{\partial}

\newcommand{\al}{\alpha}
\newcommand{\be}{\beta}
\newcommand{\ga}{\gamma}

\newcommand{\de}{\delta}
\newcommand{\De}{\Delta}

\newcommand{\ta}{\theta}
\newcommand{\Ta}{\Theta}
\newcommand{\om}{\omega}
\newcommand{\ope}{\omega_{\text{pe}}}

\newcommand{\av}{\boldsymbol{a}}
\newcommand{\bv}{\boldsymbol{b}}

\newcommand{\Vv}{\boldsymbol{V}}
\newcommand{\vv}{\boldsymbol{v}}
\newcommand{\Ev}{\boldsymbol{E}}
\newcommand{\Bv}{\boldsymbol{B}}

\newcommand{\pJ}{\mathsfbi{p}}
\newcommand{\qJ}{\mathsfbi{q}}

\newcommand{\QJ}{\mathsfbi{Q}}

\newcommand{\yv}{\boldsymbol{y}}
\newcommand{\xiv}{\boldsymbol{\xi}}
\newcommand{\TaJ}{\mathsfbi{\Ta}}

\newcommand{\OmJ}{\mathsfbi{\Omega}}

\DeclareMathOperator{\sgn}{sgn}

\newcommand{\lc}{\epsilon}
\newcommand{\ve}{\varepsilon}

\newcommand{\sg}{\sigma}
\newcommand{\lm}{\lambda}

\newcommand{\vt}{v_\text{th}}
\newcommand{\vtb}{\bar{v}_\text{th}}
\newcommand{\vtt}{\tilde{v}_\text{th}}
\newcommand{\vp}{v_\text{ph}}

\newcommand{\kc}{k_\text{char}}

\newcommand{\qe}{q_\text{even}}
\newcommand{\qo}{q_\text{odd}}

\newcommand{\dFVU}{\de_\text{FVU}}
\newcommand{\DtL}{\De t_\text{L}}

\definecolor{c0blue}{rgb}{0.12,0.47,0.71}
\definecolor{c2green}{rgb}{0.17,0.63,0.17}
\definecolor{c3red}{rgb}{0.84,0.15,0.16}
\definecolor{c7gray}{rgb}{0.5,0.5,0.5}

\shorttitle{Data-driven closures for electrostatic plasma phenomena}
\shortauthor{E.~R.~Ingelsten et al}

\title{Data-driven discovery of a heat flux closure for electrostatic plasma phenomena} 

\author{Emil R.~Ingelsten\aff{1}
	\corresp{\email{emilraa@chalmers.se}},  Madox C.~McGrae-Menge\aff{2}, E.~Paulo Alves\aff{2,3}, \and Istvan~Pusztai\aff{1}}

\affiliation{\aff{1}Department of Physics, Chalmers University of Technology, G\"{o}teborg, SE-41296, Sweden
	\aff{2}Department of Physics and Astronomy, University of California, Los Angeles, CA 90095, USA
    \aff{3}Mani L. Bhaumik Institute for Theoretical Physics, University of California at Los Angeles, Los Angeles, CA 90095, USA}

\begin{document}
    \maketitle
    


\begin{abstract}
    Progress in understanding multi-scale collisionless plasma phenomena requires employing tools which balance computational efficiency and physics fidelity. Collisionless fluid models are able to resolve spatio-temporal scales that are unfeasible with fully kinetic models. However, constructing such models requires truncating the infinite hierarchy of moment equations and supplying an appropriate closure to approximate the unresolved physics. Data-driven methods have recently begun to see increased application to this end, enabling a systematic approach to constructing closures.
    Here, we utilise sparse regression to search for heat flux closures for one-dimensional electrostatic plasma phenomena. We examine OSIRIS particle-in-cell simulation data of Landau-damped Langmuir waves and two-stream instabilities. Sparse regression consistently identifies six terms as physically relevant, together regularly accounting for more than \SI{95}{\percent} of the variation in the heat flux. We further quantify the relative importance of these terms under various circumstances and examine their dependence on parameters such as thermal speed and growth/damping rate. The results are discussed in the context of previously known collisionless closures and linear collisionless theory.
\end{abstract}
    


\section{Introduction} \label{sec:introduction}
Global modelling of multi-scale collisionless plasma phenomena is a long-standing computational challenge. Even with state-of-the-art computing resources it is often computationally infeasible to resolve the smallest scale, kinetic effects within large-scale, global domains relevant for fusion and astrophysical systems. Thus, progress in understanding these systems is bound to happen through a combination of \emph{ab initio} kinetic modelling and reduced models. In the latter category, collisionless fluid models have demonstrated their utility in global modelling \citep{Dong19,StOnge2020,Ng2020,TenBargeEvent} and as part of hybrid (kinetic-fluid) schemes \citep{Shi2021,Arzamasskiy23,Chirakkara23}. For such reduced models to be able to capture the essence of the kinetic physics at play, a systematic approach to constructing accurate fluid closures is called for. A fluid closure relates higher order fluid quantities (e.g. heat flux) -- for which the exact evolution equation equation is not retained -- to lower order fluid quantities (e.g. density, flow velocity) and fields, allowing the truncation the otherwise infinite hierarchy of fluid equations. Due to the lack of a universal closure for collisionless dynamics, each closure must be tailored to the phenomena of interest.

In collisional systems, the distribution functions of the particles remain close to local thermodynamic equilibrium, allowing for rigorous construction of closed fluid models \citep{Chapman1991,braginskii1958transport}. There is, however, no generally applicable closures for collisionless systems that are characterised by significant departures from Maxwellianity and non-local kinetic phenomena, such as wave-particle interactions. Nevertheless, there are theoretical approaches which distil various aspects of the relevant physics into the form of the closure equations. The closure by \citet{HammettPerkins} is constructed to capture Landau damping, the Chew-Goldberger-Low (CGL) closure \citep{CGL} evolves the (generally anisotropic) pressure in such a way as to conserve adiabatic invariants in collisionless magnetised plasmas, and the closure by \citet{Levermore96} is based on the principle of maximum entropy, to name a few. However, all of these approaches, while theoretically motivated, have limited scope -- such as requiring linearity or exact adiabaticity. These assumptions break down for many problems of interest, for instance in the presence of turbulence or magnetic reconnection. Furthermore, some variants are numerically difficult to work with due to spatial non-locality. In addition, there are ad-hoc closures, such as the relaxation closure that drives the pressure tensor towards an isotropic pressure \citep{Wang2015}. These have had varying success in reproducing kinetic simulation results, and may contain free parameters that cannot be determined theoretically, and the choice of which can drastically alter dynamics.

An alternative systematic line of action to obtain closures for collisionless plasma systems is to look towards data-driven methods, where the closures are constructed to conform with kinetic simulation data. Neural network-based machine learning is an effective tool to this end \citep{wangLibo20,1Maulik20,Qin2023}, though it lacks interpretability, which makes it difficult to gain intuition and generalisable understanding from. On the other hand, symbolic regression \citep{Makke2024} and sparse regression (SR) \citep{Brunton2016,Rudy2017,Schaeffer17} methods can be used to infer interpretable and generalisable equations describing a dynamical system. The models thus obtained are parsimonious, lying at the Pareto-front trading between predictive power and model complexity. SR has been used to discover the governing equations of dynamical systems in a broad range of fields previously, but it has only recently been introduced in plasma physics \citep{Dam2017,Kaptanoglu2021,Kaptanoglu2023,Alves2022}, and there is only a very limited number of attempts to use it for closure discovery. Recent work by \citet{Donaghy2023} employs SR to recover collisionless fluid equations and discover a heat flux closure in the strongly nonlinear state of a two-dimensional Harris-sheet reconnection scenario. They do not attempt to interpret the found closure, and avoid the linear regime, which is difficult due to noise at small amplitudes. Combining sparse regression and deep learning neural networks, \citet{WCheng23} recover fluid equations and the the local approximation \citep{Sharma_2006,Ng2020} of the Hammett-Perkins closure in a one-dimensional linearly Landau-damped Langmuir standing wave setup. 

Here, we employ the SINDy (Sparse Identification of Nonlinear Dynamics) algorithm \citep{Brunton2016,Rudy2017} for sparse regression to obtain a heat flux closure -- an expression for the heat flux in terms of lower order moments -- in one-dimensional (1D) electrostatic plasma scenarios. Specifically, we examine Landau-damped Langmuir waves, and in order to gain further insights into the identified closure terms and illustrate their more general nature, we also study setups exhibiting two-stream instability \citep{Stix1992} (and the following non-linear dynamics), ubiquitous in space and astrophysical systems \citep{Khotyaintsev2019}. Using particle-in-cell simulation data produced with the OSIRIS code \citep{osirisFonseca,Fonseca2008}, we search for optimally accurate expressions for the heat flux at each given model complexity (i.e. number of terms in the closure expression). Covering both linear and nonlinear stages, we follow the time evolution of the closure terms across the development of an electrostatic two-stream scenario, from growth through saturation via the formation and merging of phase-space holes. We also elaborate on the parametric dependences of the terms found, and quantify their relative importance. The expressions are interpreted in the context of the local Hammett-Perkins closure. Analytically obtained constraints between the various closure terms are also provided to support the regression results and to assist their interpretation.  

The rest of the article is organised as follows. In  \secref{sec:methods}, we describe the sparse regression method employed and the simulation setup for the systems we study, exhibiting Landau damping and growth. In \secref{sec:results}, we outline and analyse the heat flux model terms identified by SR. More specifically, we describe the results of applying SR to simulations of Landau-damped Langmuir waves and two-stream instabilities in \twosecref{sec:results_LDLW}{sec:results_TSI}, respectively. We then examine the relative importance of the various terms found in \secref{sec:results_dFVU}, relate the six terms found most consistently by SR to linear collisionless theory in \secref{sec:results_LinTh} and finally in \secref{sec:results_brakingterm} discuss a fundamentally nonlinear seventh term which is also identified as relevant by SR in many cases. We conclude by summarising our results and giving an outlook on future work in \secref{sec:conclusion}. Furthermore, we include \appref{app:LinearCollisionlessTheory}, containing a derivation of 1D electrostatic linear collisionless theory from the Vlasov-Maxwell system, as well as the constraints this imposes on the heat flux model found by SR. Finally, in \appref{app:SR_Efieldalignment} we give a demonstration of how SR works by going through how one can recover the 1D momentum equation from simulation data. 
    


\section{Methods}
\label{sec:methods}
Our aim is to find approximate, spatio-temporally local analytical expressions for the heat flux in terms of lower order fluid quantities, such that these expressions capture most of the variation in the heat flux observed in kinetic simulation data.  
When discovering heat flux closures for a given physical system, we start by performing a kinetic simulation of the system in question using the particle-in-cell (PIC) code OSIRIS. During the simulation, we export diagnostics for all fluid quantities present in the three lowest-order collisionless fluid equations (\ref{fluideqs}), namely the number density $n_\sg = \int \dd[3]{\vv} f_\sg(\vv)$, flow velocity $\Vv_{\!\!\sg} = n_\sg^{-1} \int \dd[3]{\vv} \vv f_\sg$, mass-normalised pressure tensor $\pJ_\sg = \int \dd[3]{\vv} (\vv - \Vv_{\!\!\sg})^{(2)} f_\sg$ and mass-normalised heat flux tensor $\qJ_\sg = \int \dd[3]{\vv} (\vv - \Vv_{\!\!\sg})^{(3)} f_\sg$ for each species $\sg$, as well as electromagnetic field data (electric field $\Ev$ and magnetic field $\Bv$) 
at regular time intervals. Here, $f_\sg(\vv)$ denotes the distribution function for species $\sg$, and we use notation where $\qty[\av \bv]_{ij} = a_i b_j$ and $\av^{(2)}$ is shorthand for $\av\av$. For accurate regression results it is important that the version of OSIRIS used here corrects for the otherwise occurring half time-step shifts between position and momentum data, characteristic of PIC codes using a leap-frog scheme \citep{boris1972proceedings,Hockney2021}. 
We also post-process our data to correct for staggering of the fields through linear interpolation (see \appref{app:SR_Efieldalignment} for a demonstration of the importance of correcting for such misalignments).
    


\subsection{Sparse regression}
In general, the aim of a sparse regression (SR) algorithm is to find an approximate relationship between some target quantity $y$ and a set of $M$ possibly relevant quantities $\qty{\ta_j}_{j=1}^M$, while keeping model complexity low. In the version of SINDy we use, which is one of the modified versions of the PDE-FIND algorithm described by \citet{Alves2022}, the aim is specifically to approximate $y$ as a linear combination of the $\ta_j$ quantities.  To accomplish this, we randomly select $N$ small space-time volumes from the simulation domain and integrate both $y$ and all $\ta_j$ quantities are over these small volumes to reduce noise\footnote{ 
This approach, where one reduces the effect of particle noise by integrating over the data with some kernel, is in general known as the \textit{weak formulation} of SR \citep{Schaeffer2017}. 
Integrating over small space-time volumes \say{without a kernel}, as we do, is a special case of this approach, effectively corresponding to using a space-time box function as a kernel. It should be noted that using smooth test functions, as is done in SPIDER \citep{Gurevich2024} and WSINDy \citep{Messenger2021}, provides better accuracy when higher order derivatives are important.}. We then collect the volume-integrated $y$ and $\ta_j$ quantities from all the sampled points in the domain into a vector $\yv$ and a matrix $\TaJ$ defined so that
\begin{equation}
    \qty[\yv]_i = \eval{y}_{\text{pt } i} \text{ and } \qty[\TaJ]_{ij} = \eval{\ta_j}_{\text{pt } i}.
\end{equation}
With these definitions, the task of 
approximating $y$ becomes a question of finding the coefficient vector $\xiv$ that \emph{optimally} solves the equation
\begin{equation}
    \yv = \TaJ \xiv.
\end{equation}
For us, \say{optimally} means achieving a low mean squared error with as few non-zero terms as possible,  maximising not just accuracy but also model simplicity and generalisability. Thus, our cost function looks like
\begin{equation}
    C(\xiv) = \norm{ \yv - \TaJ \xiv }^2 + \lm \norm{\xiv}_0,
\end{equation}
where $\norm{\xiv}_0$ denotes the $0$-norm of $\xiv$, i.e. the number of non-zero coefficients. The $\lm$ hyperparameter is effectively gradually increased from $0$, leading to harsher and harsher penalisation of models with a lot of non-zero terms. The end result of this procedure is a sequence of models which are optimally accurate at each given model complexity, sweeping along the Pareto front. To curb overfitting and more easily discern which terms are spurious, we perform 10-fold cross-validation -- terms in $\Ta$ which are found consistently are more likely to be physical.
The efficacy of our SR approach is demonstrated in   \appref{app:SR_Efieldalignment} through a recovery of the electron momentum equation.

In our case, we seek a heat flux closure for modelling electrostatic plasma phenomena, meaning our $y$ quantities are the elements of the heat flux tensor $\qJ_\sg$. As for the set of possibly relevant quantities $\ta_j$, in principle one would want to include all possible expressions involving $n_\sg$, $\Vv_{\!\!\sg}$, $\pJ_\sg$, $\Ev$ and $\Bv$. In practice, however, this is infeasible, since the space of possible expressions is infinite. As it turns out, even restricting to e.g. arbitrary products of the form
\begin{equation}
    \ta_j = n_\sg^\al \prod_k V_{\sg k}^{\be_k} E_{k}^{\ga_k} B_{k}^{\de_k} \prod_l p_{\sg kl}^{\ve_{kl}},
\end{equation}
where the exponents $\al$, $\be_k$, $\ga_k$, $\de_k$ and $\ve_{kl}$ are non-negative integers summing to $\leq$ some integer $s$, results in enormous term libraries $\TaJ$ even when $s$ is relatively small due to the combinatorics involved. Since having a very large term library not only increases computational cost, but also often leads to issues with convergence, choosing a term library with as few superfluous terms as possible is desired. 

For the one-dimensional electrostatic plasma problems we consider, where only electron physics is relevant over the time-scales of interest, we can first restrict ourselves to considering only $n_e$, $V_{e1}$, $E_1$, $p_{e11}$ and set $y = q_{e111}$. Since only electrons are relevant and all vectors and tensors in 1D have just a single degree of freedom, we can also suppress species and coordinate indices from now on. For convenience, we also normalise to the electron mass $m_e$, the elementary charge $e$, the speed of light $c$ and the plasma frequency $\ope=\sqrt{\bar{n}_e e^2/(\ve_0 m_e)}$ at the unperturbed electron density $\bar{n}_e$. This also normalises distances to the electron inertial length $\delta_e=c/\ope$.

To further narrow the range of possible candidate terms, we start with only those terms which are dimensionally consistent with our $y$ variable $q$, i.e. terms of the form
\begin{equation}
    \ta_j = n v_\text{th}^{\al} V^{3-\al}
\end{equation}
for some integer $\al \leq 3$, where $\vt = \sqrt{T} = \sqrt{p/n}$, defining $T = p/n$ to be (the $11$-component of) the mass-normalised temperature tensor. Inspired by the local approximation \citep{Ng2020} of the Hammett-Perkins closure, which involves a temperature gradient, we extend this initial set of candidate terms to also allow similar ones with $1^{\text{st}}$ order spatial derivatives, e.g. $n \vt \partial_x(\vt) \partial_x(V)$. It should be noted, however, that the presence of the spatial derivative in these additional terms means that the coefficient corresponding to each such term will be dimensional. For instance, the example term mentioned above with two spatial derivatives necessitates a coefficient with a dimensionality of length squared. This in turn suggests a scaling $\sim L^2$ for the coefficient in question, where $L$ is the characteristic length scale for the variation in the quantities involved.

We emphasise that restricting our term library in this way specifically is an arbitrary choice, made to limit the term library size so as to make SR convergence more likely. We start by considering dimensionally consistent terms mainly because models constructed from such terms contain only unitless coefficients, which facilitates generalisability. The restriction to integer $\al$ is made for convenience.  
Our exclusion of terms with higher-order derivatives is chiefly motivated by the fact that their inclusion would lead to difficulties with SR convergence due to the vastly increased term library size. Furthermore, closures constructed from such terms are more difficult to work with computationally, since even first-order derivatives in the expression for $\qJ$ yield second order derivatives in $\divJ{\qJ}$, and thus in the fluid equation system one needs to solve. 
Importantly, the function space we have restricted us to seems sufficient to model $q$ accurately, as we shall see. 
    


\subsection{Simulation setup}
In all of our simulations, we kinetically model an electron-proton plasma in one spatial and three\footnote{As noted above, however, we only expect the components along the single modelled spatial dimension to be of importance, meaning we are for most intents and purposes treating our simulation as 1D1V.} velocity dimensions in the centre-of-mass (CoM) frame, with physical mass ratio, a spatial resolution of $\De x = \SI{e-3}{\de_e}$ and periodic boundary conditions.
Since our simulations are all performed in the initial CoM frame and ion flow velocities remain negligibly small, every instance of an electron flow velocity $V$ below can be thought of as $V - v_\text{CoM}$, with a spatial average value of $0$. Here, $v_\text{CoM}$ is the velocity of the centre of mass. This quantity is invariant under Galilean transformations, just like $n$ and $\vt$, meaning that all terms in our term library are frame-independent with this interpretation of $V$. This is very much desirable since $q$, the quantity we are seeking to model, is a Galilean-invariant quantity\footnote{An alternative approach to ensuring Galilean or Lorentz invariance (or some other symmetry of the system) is the augmentation of simulation data through application of transformations of the corresponding type before performing SR \citep{boosts}.}.
For numerical stability, we consistently use a simulation-internal time step slightly smaller than the spatial resolution: $\SI{9.5e-4}{\ope^{-1}}$. In order to limit the amount of data output as diagnostics, we save the state of the simulation only once every 100 time steps, thus our regression analysis uses an effective temporal resolution of $\De t = \SI{0.095}{\ope^{-1}}$.
    


\subsubsection{Landau-damped Langmuir waves}
When studying Landau-damped Langmuir waves, we initialise the plasma as a Maxwell distribution with various non-relativistic thermal speeds $\vt \sim \SI{0.01}{c}$ using a domain size of $\SI{0.256}{\de_e}$ with $10^5$ $\qty(10^4)$ electrons (ions) per cell. To excite Langmuir waves, we then apply and smoothly turn off an external sinusoidal $\Ev$-field perturbation propagating in the $+x$ or $-x$ direction with wavenumber $\abs{k} = 4\pi / \qty(\SI{0.256}{\de_e})$ and a frequency $\om_r$ matching that of the analytic Langmuir mode. More specifically, this is done by using a single-cycle sine squared envelope, reaching maximum amplitude at $\ope t = 3$ and being fully turned off at $\ope t = 6$. The values of $\vt$ considered, along with corresponding $\abs{k} \lm_{D,e}$ values (where $\lm_{D,e}$ is the electron Debye length), as well as frequencies and growth rates of the resulting Langmuir waves, are shown in \tabref{tab:pmVals_Ldamping}.

\begin{table}
    \centering
    \begin{tabular}{ccccccc}
        \toprule
        Initial $\vt/(\SI{0.01}{c})$: & 0.8 & 0.9 & 1 & 1.1 & 1.25 & 1.5 \\
        \midrule
        Resulting $\abs{k} \lm_{D,e}$: & 0.393 & 0.442 & 0.491 & 0.540 & 0.614 & 0.736 \\
        \midrule
        $\om_r/\ope$ ($\pm 0.06$*): & 1.24 & 1.31 & 1.36 & 1.42 & 1.50 & 1.67 \\  
        \midrule
        Resulting $\om_r/\qty(k\vt)$: & 3.16 & 2.97 & 2.77 & 2.63 & 2.44 & 2.27 \\ 
        \midrule
        $\ga/\ope$: & $-0.0692$ & $-0.109$ & $-0.150$ & $-0.195$ & $-0.291$ & $-0.429$ \\
        \midrule
        $\ope \DtL$: & 15.0 & 15.0 & 15.0 & 14.5 & 8.3 & 6.4 \\
        \midrule
        $\ope t_b$: & 7.30 & 7.66 & 8.10 & 8.61 & 9.56 & 11.3 \\ 
        \bottomrule
    \end{tabular}
    \caption{The values of $\vt$ used when studying Landau-damped Langmuir waves, the frequency $\om_r$ and the (negative) growth rate $\ga$ of the excited Langmuir wave and the duration $\DtL$ of the period of exponential decay, over which SR is applied, as well as the estimated bounce time $t_b$ for trapped electrons in each case. We also list the values of $\abs{k} \lm_{D,e}$ and $\om_r/\qty(k\vt)$ resulting from the other parameters. The frequencies $\om_r$ are calculated via Jacobsen interpolation \citep{Jacobsen2007} of the peaks in the Discrete Fourier Transform (DFT) spectrum for the $\Ev$-field, with uncertainty (*) corresponding to half the DFT bin size. The growth rate $\ga$ is calculated via linear regression on logarithmised data of the average $\Ev$-field energy density over the period of exponential decay. The estimated bounce time is calculated as $t_b = \sqrt{ m_e / \qty( e \abs{k} E_\text{RMS} ) }$, where $E_\text{RMS} = \sqrt{\expval{E^2}}$ is the spatial RMS average of the $\Ev$-field magnitude at the point in time when the external drive is switched off.
    \label{tab:pmVals_Ldamping}
    }
\end{table}

\begin{figure}
    \centering
    \includegraphics[width=\textwidth]{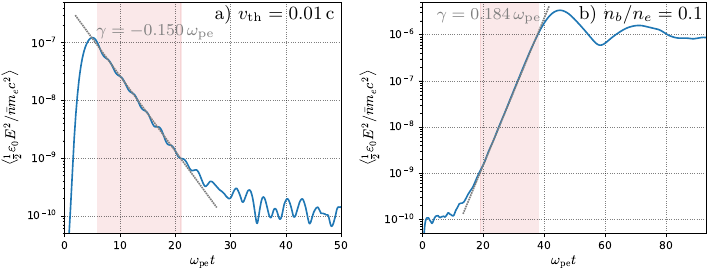}
    \caption{The evolution of the spatially averaged $\Ev$-field energy density $\expval{\frac{1}{2} \ve_0 E^2}$ over time a) in the Landau damping case where $\vt = \SI{0.01}{c}$ and b) in the two-stream case where $n_b/n_e = 0.1$, normalised to the rest energy of an electron and the unperturbed electron number density $\bar{n}$. The linear decay/growth phase is highlighted in red.}
    \label{fig:logEsq}
\end{figure}

After the external forcing is removed, the system is left to evolve self-consistently, with the resulting Langmuir waves decaying due to Landau damping -- initially exponentially, with only \emph{linear} processes involved, as can be seen in \figref{fig:logEsq}a. The PDE-FIND algorithm is then applied to find a closure for $q$ during the timeframe of length $\DtL$ where decay is judged to be exponential (e.g. $6.0 < \ope t < 21.0$ for initial $\vt \leq \SI{0.01}{c}$ -- see also \figref{fig:logEsq}a, where this time range is highlighted in red). In total, $\sim \SI{6}{\percent}$ of this space-time range is randomly sampled per cross-validation fold, of which there are 10. The values of $\DtL$ for all values of $\vt$ considered are listed in \tabref{tab:pmVals_Ldamping}, together with the estimated bounce times $t_b$ for trapped electrons. Note that for the four lower thermal speeds considered, decay is exponential for roughly twice the bounce time, meaning that we may expect slight nonlinear trapping effects towards the end of the sampled time window. At higher values of $\vt$, however, these effects are drowned out by numerical noise present at the low perturbation amplitudes reached towards the end of exponential decay. Indeed, this is the reason for $\DtL$ being lower than $t_b$ for the two highest $\vt$ values considered -- in these cases, the decay is so rapid that the perturbation disappears in the numerical noise before trapping effects become visible.

\subsubsection{Two-stream instabilities}
Apart from studying Landau damping, we also examine a setup exhibiting Landau growth, namely a two-stream unstable plasma. In this case, we choose a larger domain size of $\SI{2.048}{\de_e}$ to limit the effects of the periodic boundary conditions employed. We initialise the ions in equilibrium, with the electrons split into counterflowing equal temperature Maxwellian populations -- a \textit{core} population with density $n_c$ and flow velocity $V_c$, and a \textit{beam} population with density $n_b$ and flow velocity $V_b$. The thermal speed for each population individually is $u_\text{th} = \SI{3.16e-3}{c}$. We vary $n_b$ as a fraction of the total electron density $n_e$, keeping the relative velocity $V_\text{rel} = V_c - V_b$ constant at $\SI{0.02}{c}$ and staying in the zero-current frame by enforcing $n_c V_c + n_b V_b = 0$. Specifically, we consider the range of values for $n_b/n_e$ listed in \tabref{tab:pmVals_2stream}. In these simulations, we use \SI{2e4}{} (\SI{200}{}) electrons (ions) per cell. 

Note that the non-zero relative velocity $V_\text{rel}$ between the populations means that the combined electron population had a thermal speed $\vt = \sqrt{p/n} > u_\text{th}$. More specifically, this combined thermal speed is $\vt = [ u_\text{th}^2 + (n_b/n_e) \qty( 1 - n_b/n_e ) V_\text{rel}^2 ]^{1/2}$,
with a maximum of $\vt \approx \SI{1.05e-2}{c}$ for $n_b/n_e = 0.5$.

\begin{table}
    \centering
    \begin{tabular}{cccccccc}
        \toprule
        $n_b / n_e$: & 0.01 & 0.02 & 0.05 & 0.1 & 0.2 & 0.4 & 0.5 \\
        \midrule
        $\vt / (\SI{0.01}{c})$: & 0.373 & 0.422 & 0.503 & 0.678 & 0.860 & 1.03 & 1.05 \\
        \midrule
        $\de_e \kc$: & -60.8 & -59.9 & -60.2 & -62.3 & -64.3 & -67.6 & -67.4 \\ 
        \midrule
        Resulting $\abs{\kc} \lm_{D,e}$: & 0.227 & 0.253 & 0.303 & 0.422 & 0.553 & 0.696 & 0.708 \\
        \midrule
        $\om_r/\ope$ ($\pm 0.03$*): & 1.00 & 0.96 & 0.88 & 0.80 & 0.66 & 0.19 & 0 \\ 
        \midrule
        $\ga/\ope$: & 0.0515 & 0.0837 & 0.137 & 0.184 & 0.234 & 0.273 & 0.276 \\ 
        \midrule
        $\ope \DtL$: & 31.3 & 23.7 & 20.9 & 19.0 & 13.3 & 12.3 & 11.9 \\ 
        \midrule
        $\ope t_b$: & 5.95 & 4.95 & 3.54 & 2.94 & 2.54 & 2.28 & 2.29 \\ 
        \bottomrule
    \end{tabular}
    \caption{The values of $n_b/n_e$ examined when studying two-stream instabilities and the corresponding values of $\vt$, as well as the frequency $\om_r$, growth rate $\ga$ and characteristic wavenumber $\kc$ for the excited perturbation (also listing $\abs{\kc} \lm_{D,e}$), together with the duration $\DtL$ of exponential growth and the bounce time $t_b$. The same methods are used to calculate $\om_r$ and $\ga$ as in the Landau damping cases, described in \tabref{tab:pmVals_Ldamping}. 
    The values of $\kc$ are calculated by using a weighted average over the DFT spectrum at the end of linear growth, using the absolute value of the Fourier amplitude squared as the weighting. The minus signs signify propagation towards $-x$.
    Similarly to what was done in the Landau damping cases, we estimate $t_b = \sqrt{ m_e / \qty( e \abs{\kc} E_\text{RMS} ) }$, with both $\kc$ and $E_\text{RMS}$ in this case being evaluated at the time of maximum average $\Ev$-field energy density. Note that while $\kc$ at this time is slightly different than the values listed in this table, the difference is marginal ($\sim \SI{1}{\percent}$).
    \label{tab:pmVals_2stream}
    }
\end{table}

The counterflowing electron populations drive wave growth via inverse Landau damping. Similarly to the decay of the Langmuir waves above, this growth is initially exponential. It eventually saturates, however (as can be seen in \figref{fig:logEsq}b), leading to the formation of phase-space electron holes. With the data from these simulations, SR is performed (a) over the linear part of the growth phase, and (b) over small time slices of length $\SI{1.9}{\ope^{-1}}$ covering both the growth phase and the saturated phase to study how the closure coefficients evolve over time, as illustrated in \figref{fig:twostream} -- in both cases for all values of $n_b/n_e$ listed in \tabref{tab:pmVals_2stream}. Part (a) here is very much analogous to what was done for the Landau damping case. For example, in the case where $n_b/n_e = 0.1$, the time range sampled is $19.0 < \ope t < 38.0$, highlighted in red in \figref{fig:logEsq}b, meaning $\ope \DtL = 19.0$. In part (b), the time slices are centred on time steps $t = 1.9\times\qty{1, 2, 3, \ldots}\ope^{-1}$, covering the entire simulation domain up to the last such time step which is $> 0.95 \ope^{-1}$ from the end of the simulation, so that every time slice falls entirely within the domain of the simulation. In both of these cases, each of the 10 cross-validation folds randomly samples $\sim\SI{2.5}{\percent}$ of the data in the space-time ranges of interest.

Since we are now dealing with exponential growth rather than exponential decay, the influence of noise towards the end of the linear part of the process is significantly decreased compared to the situation in the Landau damping simulations. This is also clearly visible in the stronger relationship between $\DtL$ and $t_b$ in these simulations -- we consistently have $\DtL \sim 5.5 t_b$, meaning growth is exponential for a little over five times the bounce time.
This suggests that there is a high probability of nonlinear trapping-related effects being present to some extent in the data towards the latter half of the sampled time range.

The excited perturbations in this case are more broad-spectrum than the Langmuir waves examined above, necessitating the introduction of characteristic wavenumbers $\kc$, calculated as outlined in the caption of \tabref{tab:pmVals_2stream}. The temporal spectrum is dominated by a single frequency peak however, at the value of $\om_r$ listed in \tabref{tab:pmVals_2stream}. Like in \tabref{tab:pmVals_Ldamping}, we also for convenience show $\abs{\kc} \lm_{D,e}$.

For simplicity and to more easily compare our results with those from the Landau-damping case, we only consider the combined electron species rather than treating the counter-streaming populations separately when performing SR. That is, all of our closure models are models of the total electron heat flux, and our term library is constructed from fluid quantities relating to the entire electron population.
We note, however, that for the purposes of modelling two-stream unstable systems in fluid codes it is likely more practical to treat the two electron populations as separate species, with their own respective closures. Preliminary results suggest that applying SR when using such an approach also yields broadly similar closure terms as those identified here, in appropriately chosen reference frames. 
To avoid diverting our focus, we leave a more detailed analysis of two-stream instability along these lines, with separate closures for the beam and core populations, outside the scope of this article.   
    


\section{Results}
\label{sec:results}
For both of the setups we considered, SR yields very similar results. In both cases, a 6-term model $q = \qe + \qo$ was found, where
\begin{equation}
\begin{cases}
    \qe = A_1 n \vt^2 V + A_2 \vt^3 \pd_x n + A_3 n \vt^2 \pd_x \vt \\
    \qo = A_4 + A_5 n \vt^3 + A_6 n \vt^2 \pd_x V.
\end{cases}
\label{eq:qmodel_6term}
\end{equation}
The split into $\qe$ and $\qo$ is based on the dependence on the propagation of the perturbations involved. While the coefficients in front of the $\qe$ terms are independent of propagation direction (and thus \say{even in $k$}), the $\qo$ coefficients switch sign if the propagation direction is reversed (and are thus \say{odd in $k$}). This also means that if there is no wave propagation, or when oppositely propagating waves are of similar amplitudes, such as in a standing-wave scenario, all $\qo$ coefficients go to zero. In several cases, an additional term $\propto n \vt V^2$ is found. As this term mostly appears for low $\abs{\ga}$ -- specifically towards the end of linear growth or decay processes when $\abs{\ga}$ is decreasing, it appears to help capture weak nonlinear trapping effects.

We note that, due to the presence of spatial derivatives in terms 2, 3 and 6, the corresponding coefficients are dimensional, having units of length. When plotting them, they are implicitly given in units of $\de_e$, though often re-scaled by a factor of $10^2$ for readability. In other words, a label $A_2 \times \SI{e2}{}$ should read $10^2 A_2 / \de_e$. Similarly, $A_4$ has the units of mass-normalised heat flux, or number density times velocity cubed, and is implicitly given in units of $\bar{n} c^3$. Thus, a curve labelled $A_4 \times \SI{e6}{}$ shows $10^6 A_4 / (\bar{n} c^3)$.

We consistently quantify the error of the various models found by SR using the fraction of variance unexplained (FVU), defined as $\text{FVU} = \sum_i (y_i-\hat{y}_i)^2/\sum_i(y_i-\overline{y})^2$, where $y_i$ is the value of the $y$ quantity at the $i^{\rm th}$ sampling point, $\hat{y}_i$ is the $y$-value predicted by the model at that point, $\overline{y} = \frac{1}{N} \sum_i y_i$ is the mean $y$-value, and the sums run over the $N$ samples. As stated in \secref{sec:methods}, the $y$ quantity of interest to us is the total electron heat flux $q = q_{ e 111 }$.
    


\subsection{Landau-damped Langmuir waves} \label{sec:results_LDLW}
In the simulations of Landau-damped Langmuir waves, SR found the six-term closure in \eqref{eq:qmodel_6term} consistently, with an FVU of 2-\SI{7}{\percent}. The FVU increases with higher initial $\vt$, and corresponding stronger damping. In simulations of standing waves (i.e. the sum of oppositely propagating waves), only the $q_\text{even}$ terms are found, consistent with the lack of a preferred direction.

\begin{figure}
    \centering
    \includegraphics[width=\textwidth]{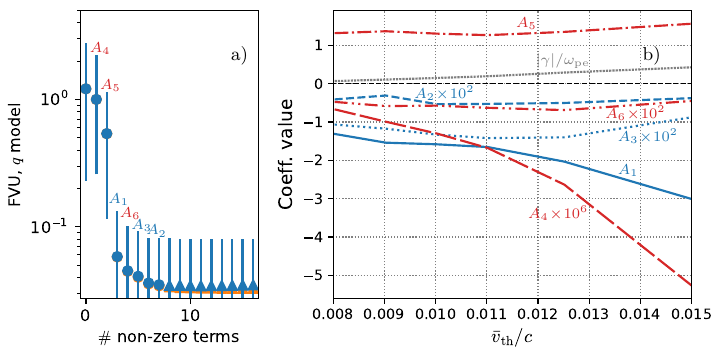}
    \caption{Results from the Landau damping simulations: a) FVU of the successive closures found for $\vt/c = 0.01$ (where $\ga = -0.150$), with each new term indicated by its coefficient in \eqref{eq:qmodel_6term}. Circles, unlike triangles, denote consistently found terms, and blue (orange) marker color corresponds to performance on testing (training) data. Note that while FVU is typically smaller than unity, it is higher than unity in certain situations. For example, this is the case for the $0$-term model, which is identically zero, meaning $\hat{y}_i = 0$ for all $i$, while the mean of the $y$ data is $\overline{y} \ne 0$.  b) The dependence of the closure coefficients on $\vt$, for a wave propagating in the $+x$ direction. Blue (red) lines/symbols indicate $q_\text{even}$ ($q_\text{odd}$) coefficients. For comparison, $\abs{\ga} / \ope$ is plotted as a gray tightly dotted line.}
    \label{fig:landaudamping}
\end{figure}

We note that in some cases, additional terms are sometimes found consistently. For example, the unannotated seventh circle marker in \figref{fig:landaudamping}a corresponds to a term $\propto n V^2 \pd_x V$. In particular, at low $\abs{\ga}$, where nonlinear trapping effects are expected to be slightly more important, a term $\propto n \vt V^2$ is found, likely helping capture these weak nonlinear effects.

Since we are examining an electrostatic 1D setting affected by Landau damping, one might expect the closure to be similar to the local approximation \citep{Sharma_2006,Ng2020} of the Hammett-Perkins closure \citep{HammettPerkins}: $q\sim -\qty(\chi/\abs{\kc}) n\vt^2\partial_x \vt$, where $\kc$ is a characteristic wavenumber. The value one should choose for the heat diffusivity $\chi$ depends on which values of $\om/(k\vt)$ are of interest, with the original Hammett-Perkins paper focusing on regimes relevant for the ion temperature gradient (ITG) instability. In our closure, the $A_3$ term can be identified as playing this role. 
Sparse regression finds $\vt$-dependent coefficients, see \figref{fig:landaudamping}b. For most coefficients, however, this $\vt$-dependence is significantly weaker than the $\vt$-dependence of $\ga$ over the examined range of values---the major exception being the constant term $A_4$, which does not affect $\pd_x q$, the quantity we are ultimately in need of a closure for. There is one coefficient with non-negligible $\vt$-dependence which \textit{does} affect $\pd_x q$, however---namely, $A_1$. 
Interestingly, we find that the improvement in predictive power caused by the introduction of this $A_1 n v_{\text{th}}^2 V$ term into the model is significantly larger than the one coming from the Hammett-Perkins-like $A_3 n v_{\text{th}}^2 \pd_x \vt$ term, as seen in \figref{fig:landaudamping}a.
    


\subsection{Two-stream instabilities}
\label{sec:results_TSI}
\begin{figure}
    \centering
    \includegraphics[width=\textwidth]{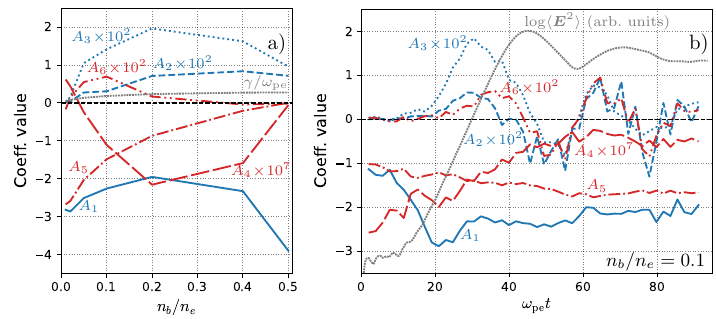}
    \caption{Results from the two-stream instability simulations: a) The dependence of the growth phase closure coefficients on $n_b / n_e$. For comparison, $\ga / \ope$ is plotted as a gray tightly dotted line. b) The evolution of the coefficients over time in the case where $n_b / n_e = 0.1$. Blue (red) lines indicate $q_\text{even}$ ($q_\text{odd}$) coefficients. Here, the gray tightly dotted line is the logarithm of the spatially averaged $\Ev$-field energy density in arbitrary units, recognisable from \figref{fig:logEsq}b.}
    \label{fig:twostream}
\end{figure}

As noted above, broadly speaking the same 6-term model is found for two-stream instabilities as for Landau-damped Langmuir waves, at an even lower FVU than in the Landau damping simulations of between \SI{0.5}{\percent} and \SI{5}{\percent}. Furthermore, the term $\propto n \vt V^2$ is again found in cases with low $\abs{\ga}$ -- in fact, it appears for all $n_b/n_e < 0.4$ (but is left out of \figref{fig:twostream} for readability). Also, some coefficients are close to zero at certain $n_b/n_e$ values -- see \figref{fig:twostream}a. Such near-zero coefficients are generally not found consistently. For example, when $n_b = 0.5 n_e = n_c$, the lack of a preferred direction forces all $q_\text{odd}$ coefficients to zero, just like for a Landau-damped standing wave. The sub-\SI{1}{\percent} FVU values are achieved in the middle of the examined beam density range, where amplitudes are relatively small but growth is still rapid enough for trapping effects to be mostly negligible throughout the linear growth phase.

Both in these setups and the Landau damping ones examined above, the growth (or decay) phase value of $A_3$ tends to be roughly twice $A_2$. An exact factor of 2 would correspond to having a single term $\propto \vt \partial_x p = v_\text{th}^3 \partial_x n + 2 n v_\text{th}^2 \partial_x \vt$, i.e. a pressure gradient driven contribution to the heat flux, whereas the $A_2$ and $A_3$ terms on their own correspond to contributions due to density and temperature gradient driven heat flux contributions, respectively. Notably, while $A_{2}$ and $A_3$ are negative for Landau damping they are positive in the growth phase of the instability, where inverse Landau damping occurs. Interestingly, the same holds for $A_6$ in most cases examined here, despite the fact that the wave propagation is towards $-x$ here, whereas it is towards $+x$ for the Langmuir waves examined in \figref{fig:landaudamping}. As we will see in \secref{sec:results_LinTh}, this can be explained quite well by the constraints imposed on the coefficients from linear collisionless theory. Being a $k$-odd term, the $A_6$ term represents a contribution to the heat flux coming from the pressure times the rate of change of flow velocity in the direction of local wave propagation. 

\begin{figure}
    \centering
    \includegraphics[width=\textwidth]{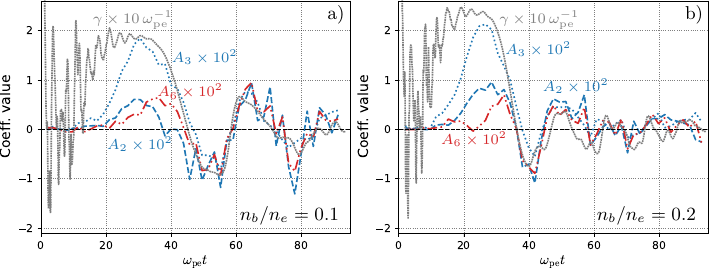}
    \caption{The three growth/damping-related coefficients $A_2$, $A_3$ and $A_6$, compared with the instantaneous growth rate, a) in the case where $n_b/n_e = 0.1$ also shown in \figref{fig:twostream}b and b) in the case where $n_b/n_e = 0.2$. As we can see, $A_3$ is very nearly precisely proportional to $\ga$, while the other two coefficients are also strongly correlated with it.}
    \label{fig:A236_vs_gamma}
\end{figure}

So far we have considered model coefficients obtained only for the growth phase of the instability. Now we will consider both the growth phase and the saturated phase of the simulation, with coefficients obtained in a time-resolved manner. We find that while the same model terms are sufficient to accurately approximate the heat flux throughout the simulation\footnote{This is not necessarily what one would expect. In general, the model terms -- or even the modelling approach -- might need to be adapted to the various phases of the system's evolution. To inform one's choice of model, and to quantify the complexity of the training data, which affects the difficulty of recovering/discovering terms, the Shannon information entropy metric \citep{Vasey2025,Kaptanoglu2023NL} can be used.}, some of the coefficients vary significantly across these phases. Overall, $A_3$ is very well correlated with the instantaneous growth rate, as is $A_2$ and $A_6$. This is seen in \figref{fig:twostream}b, but is even more apparent if one plots the three coefficients in question against the instantaneous growth rate $\ga(t) = \frac{1}{2} \pd_t \ln \expval{\Ev^2}$ itself -- see \figref{fig:A236_vs_gamma}. What is interesting is that while the amplitude of the oscillation in $A_3$ decreases along with that of the oscillation in $\ga(t)$, the other two growth-related coefficients exhibit a far smaller change in oscillation amplitude from the growth phase to the saturated phase. It is also notable that while $A_2$ and $A_6$ are practically equal during the saturated phase, they are desynchronised during the growth phase, with $A_2$ reaching its growth-phase maximum earlier than $A_6$. Note, however, that while $A_2$ and $A_6$ are out of phase in this way during the growth phase for all examined values of $n_b/n_e$ (disregarding the symmetric setup with $n_b/n_e = 0.5$, where $A_6 = 0$), the exact nature of their relationship varies depending on $n_b/n_e$, as seen comparing the two cases shown in \figref{fig:A236_vs_gamma}. In general, they synchronise earlier in simulations with higher beam density. Furthermore, the fact that the oscillation amplitudes of $A_2$ and $A_6$ are approximately equal only holds when $n_b/n_e \sim 0.1$, as one might suspect from the fact that $A_6 \to 0$ as $n_b/n_e \to 0.5$ by virtue of being $k$-odd.

Compared to the growth-related terms, $A_1$ and $A_5$ (as well as $A_4$) vary relatively slowly over time -- especially $A_5$. Overall, the value of $A_1$ fits quite well with the heuristic $A_1 \sim -3 + \frac{1}{2} \sgn(k) A_5$ obtained from linear theory in the limit where $\abs{\ga} \ll \om_r \sim \ope \sim \abs{k}\vtb$ (see \secref{sec:results_LinTh}), the actual values in this case being $\abs{\ga} < \SI{0.2}{\ope}$, $\om_r = \SI{0.80}{\ope}$ and $\abs{k}\vtb \sim \SI{0.4}{\ope}$. The one notable exception to this is the very early growth phase, where the prevalence of high-$k$ noise means that the physics involved is very far from this limit. In fact, in the later parts of the saturated phase, the heuristic performs better than expected, given the fact that the merging of electron holes should decrease the characteristic wave number $\abs{k}$ as time goes on. 
There also appears to be a contribution from the growth rate $\ga$ on top of this, leading to slight oscillations in $A_1$ over time, consistent with the constraints imposed by linear theory as outlined in \secref{sec:results_LinTh}.

When it comes to giving a physical interpretation of these non-growth-correlated terms, $A_4$ can be thought of as a global heat flux induced by passing waves. As seen in \secref{sec:results_LinTh}, this term is beyond the purview of linear theory -- in fact, the exact mechanism giving rise to this contribution is unclear. However, the $A_4$ term does not affect the divergence of the heat flux, which is what one is ultimately seeking to model when creating a closure for the pressure equation. The $A_1$ term is transparently a product of pressure and flow velocity, and it provides either the most or the second most important contribution to $q$, depending on $n_b/n_e$. 
Its appearance in our closure might be related to the fact that $\qty{\Vv\pJ}$ is part of the expression relating the energy flux $\QJ = \int \dd[3]{\vv} \vv^{(3)} f$ to the heat flux $\qJ$: $\QJ = n \Vv^{(3)} + 3\qty{\Vv\pJ} + \qJ$. Thus, having a term in our $q$ model equal to specifically $-3\qty{\Vv\pJ}$ (i.e. having $A_1 = -3$, which is quite a typical value found by SR) would signify a cancelling of this term in the energy flux. In the 1D pressure equation, having an $A_1$ term as part of $q$ similarly leads to partial cancellations. Specifically, writing the rest of $q$ (i.e. $q$ excluding the $A_1$ term) as $q_r$, the equation reduces to 
\begin{equation}
    \pd_t p + \qty(1 + A_1) V \pd_x p + \qty(3 + A_1) p \pd_x V + \pd_x q_r = 0,
\end{equation}
so that the value $A_1 = -1$ would cancel the second term and $A_1 = -3$ would cancel the third term. 
The final of the three most important terms in our $q$ model (see \secref{sec:results_dFVU}), i.e. the $A_5$ term, 
is $k$-odd. Thus, it is clearly related to the wave propagation direction. 
Furthermore, in setups like ours which are in the CoM frame, $V$ oscillates around zero while $n$ and $\vt$ have positive equilibrium values. As discussed in \secref{sec:results_LinTh}, this means that all of the terms in our 6-term model contribute to $q$ at first order in perturbation theory. While there are many terms in our term library that do this, there are actually only two which contribute to $q$ at zeroth order in perturbation theory due to how our term library is constructed -- precisely the $A_4$ and $A_5$ terms. This may be related to why they are identified by SR as being useful for modelling $q$.

The fact that some of the coefficients correlate with growth (or decay) -- and as such differ significantly between the growth and the saturated phases of the instability -- means that we should not expect to find a closure with fixed coefficients which is accurate throughout both phases \emph{with respect to the contribution from these terms}. To capture both phases, one would need coefficients which are informed about the phase, through e.g. a volume-averaged electric-to-thermal energy ratio.
We do not aim to provide such closures here. However, as we shall see in the next section, the growth-related terms contribute relatively little to the accuracy of the model compared to the $A_1$, $A_4$ and $A_5$ terms, meaning the closure we obtain in the growth phase remains quite accurate even in the saturated phase.



\subsection{Quantifying the importance of terms: $\dFVU$}
\label{sec:results_dFVU}
\begin{figure}
    \centering
    \includegraphics[width=0.6\textwidth]{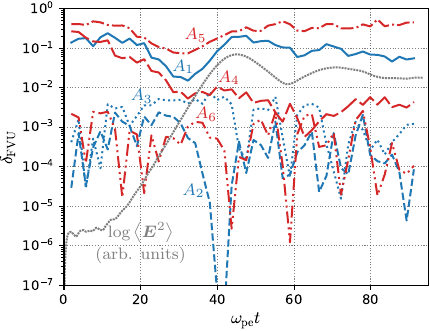}
    \caption{The importance of the six terms found by SR as they vary over time in the two-stream unstable setup with $n_b/n_e = 0.1$, as measured by $\dFVU$.}
    \label{fig:deltaFVU_tDep}
\end{figure}

To get a better sense of the circumstances under which each term in our closure is important, let us quantify their individual contributions by how much their exclusion increases the FVU of the closure, a measure we will refer to as $\dFVU$. Doing this for the various time slices examined in \figref{fig:twostream}b, we get \figref{fig:deltaFVU_tDep}.

The two by far most important terms are the two terms with order unity coefficients, i.e. $A_5$ and $A_1$. The next most important term by $\dFVU$ is the constant term $A_4$, although as noted previously, this term is not very relevant to the accuracy of the closure since it has no impact on $\pd_x q$. Among the growth-related terms, $A_3$ is overall the most important by some margin, while $A_6$ and $A_2$ are the least important terms -- $A_6$ in general being slightly more important than $A_2$ in this case. Interestingly, $A_2$ is important mostly in the first half of a linear growth or decay process, while $A_6$ mostly matters during the latter half. 
This agrees very well with what one might guess from solely looking at the sizes of the coefficients in question in \figref{fig:twostream}b. While it is not obvious that this should be the case, it is reasonable from the perspective that if the best fit for a coefficient is zero at some point in time, its importance is necessarily also zero. Since the achieved FVU is at best $\sim \SI{5e-3}{}$, any term with $\dFVU \lesssim \SI{e-4}{}$ can be safely assumed to be irrelevant at our level of accuracy for describing the physics during that time range.

\begin{figure}
    \centering
    \includegraphics[width=\textwidth]{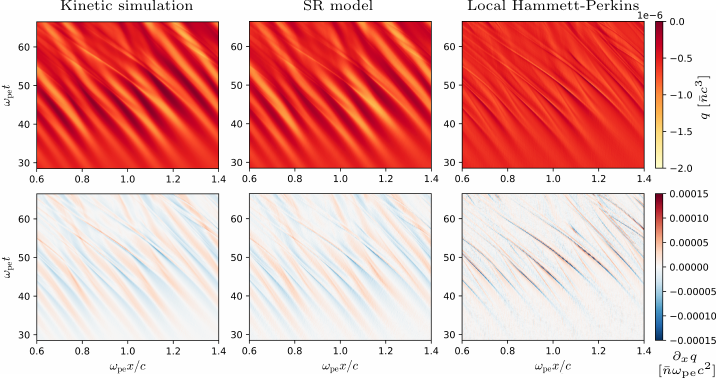}
    \caption{A comparison of the OSIRIS $q$ data (top left) from the $n_b/n_e = 0.1$ simulation with our 6-term SR model (top middle) and a local Hammett-Perkins model equivalent to keeping only $A_3$ and $A_4$ (top right), as well as the resulting $\pd_x q$ (bottom row). Both models are trained solely on data from the linear growth phase, corresponding to $19.0 < \ope t < 38.0$. The 6-term model performs very well even in the saturated regime, corresponding to $\ope t \gtrsim 40$. All mass-normalised heat fluxes, like the $A_4$ coefficient, are given in units of $\bar{n}c^{3}$, and spatial derivatives of such quantities are given in units of $\bar{n}\ope c^{2}$.}
    \label{fig:qSR_vs_qSim}
\end{figure}

The fact that the most important terms, $A_1$ and $A_5$, vary quite slowly means that regardless of which time range we perform the regression on, we should expect the resulting closure to work quite well over the entire simulated time range. And indeed, this is what one finds if one plots the 6-term model $q = \qe + \qo$ with coefficient values from (e.g.) the growth phase over the entire spacetime domain of the simulation and compares it to a plot of the actual $q$ output by OSIRIS -- see \figref{fig:qSR_vs_qSim}.

This is especially promising since one of the primary use cases envisioned for these kinds of closures is sub-grid scale modelling within a larger simulation, where the instability occurs on a very short time scale compared to the overall time scales of interest. In such a situation, modelling the saturated phase \say{end state} where $\ga \approx 0$ is the most important. The fact that there is no growth on average means that the 6-term model can be reduced to a three-term model with only $A_1$, $A_4$ and $A_5$ -- and of course, providing a value for $A_4$ is unnecessary if one is only interested in solving the fluid equations, since $A_4$ does not affect $\pd_x q$.

In general, using a 6-term model trained solely on growth phase data like in \figref{fig:qSR_vs_qSim} tends to yield FVU $\sim$ \SI{1}{\percent} during the growth phase and FVU $\sim$ 5-\SI{10}{\percent} in the saturated phase, while using a 6-term model trained on data from the saturated phase where there is no net wave growth (or, more-or-less equivalently, a 3-term model with only the $A_1$, $A_4$ and $A_5$ terms) yields FVU $\sim$ 5-\SI{10}{\percent} in the growth phase and FVU $\sim$ 2-\SI{5}{\percent} in the saturated phase. Thus, our 6-term (or indeed 3-term) model seems to be largely sufficient for modelling the saturated phase, despite the presence of nonlinear phenomena like particle trapping and soliton-like phase-space electron holes.

\subsection{Comparison with constraints from linear theory} \label{sec:results_LinTh}
During linear decay or growth, the plasma should be well described by linear collisionless theory. As explained in \appref{app:LinearCollisionlessTheory}, this gives us two predictions relating the different closure coefficients, namely
\begin{equation}
\begin{dcases}
    k A_6 = \be + \qty[ -k A_2 \om_r + \frac{1}{2} kA_3 \qty(1 + \Phi_-) \om_r - \frac{1}{2} A_5 \qty( 1 + 3 \Phi_+ ) \ga ] \frac{k\vtb}{\abs{\om}^2} \\
    A_1 = -3 - \al + \qty[ -k A_2 \ga + \frac{1}{2} k A_3 \qty( 1 + \Phi_+ ) \ga + \frac{1}{2} A_5 \qty( 1 + 3 \Phi_- ) \om_r ] \frac{k\vtb}{\abs{\om}^2},
    \label{eq:A1A6LinThPred}
\end{dcases}
\end{equation}
where we are using shorthand notation 
\begin{equation}
    \al = \frac{\ope^2 + \ga^2 - \om_r^2}{k^2\vtb^2}, \quad \be = \frac{2\om_r\ga}{k^2\vtb^2}, \quad \Phi_\pm = \frac{\ope^2 \pm \abs{\om}^2}{k^2\vtb^2}.
\end{equation}
Let us first consider some limiting cases. First, take
a marginally stable perturbation with $\ga \to 0$, where \eqref{eq:A1A6LinThPred} simplifies to
\begin{equation}
\begin{dcases}
    A_6 = \qty[ -A_2 + \frac{1}{2} A_3 \qty(1 + \frac{\ope^2 - \om^2}{k^2\vtb^2}) ] \frac{k\vtb}{\om} \\
    A_1 = -3 - \frac{\ope^2 - \om^2}{k^2\vtb^2} + \frac{1}{2} A_5 \qty( 1 + 3 \frac{\ope^2 - \om^2}{k^2\vtb^2} ) \frac{k\vtb}{\om},
\end{dcases}
\label{eq:A1A6LinThPred_gammato0}
\end{equation}
with $\om = \om_r$ real. We can immediately see that one solution of the first of these equations is $A_2 = A_3 = A_6 = 0$, in agreement with the relationship $A_{2,3,6} \sim \ga$ found via SR. The second equation is less straightforward to interpret. If $\om \sim \ope \sim \abs{k}\vtb$, like in most of our simulations, we expect $A_1 \sim -3 + \frac{1}{2} \sgn(k) A_5$ -- and this should hold as a rule of thumb even when $\ga$ is non-zero but small compared to $\om_r$. Qualitatively, this agrees decently with our results, even those from the growth phase where $\ga > 0$ (but in most cases $< \om_r$). Generally, both $A_1$ and $A_5$ are order unity, and $A_1$ is shifted a bit upwards from $-3$ in \twofigref{fig:landaudamping}{fig:twostream} for all setups we consider except the symmetric two-stream setup, matching the fact that $\sgn(k A_5) = +1$. That this case should disagree with our rule of thumb is not surprising, since it has $\om_r \approx 0$, while $\ga$ is finite.

\begin{figure}
    \centering
    \includegraphics[width=\textwidth]{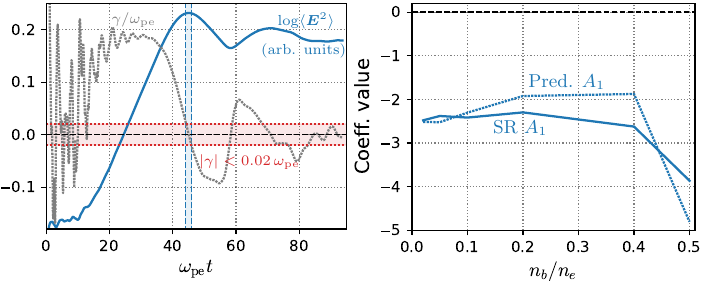}
    \caption{a) The time range around peak $\Ev$-field energy density where $\abs{\ga}/\ope < 0.02$ for $n_b/n_e = 0.1$. b) The values of $A_1$ found by SR in the two-stream simulations during the equivalent time range for all examined values of $n_b/n_e$, compared to the linear theory prediction at $\ga = 0$ given by \eqref{eq:A1A6LinThPred_gammato0}, inserting the values of $A_5$ found by SR.}
    \label{fig:A1LinThComp_gammato0}
\end{figure}

The growth rate is truly negligible mainly during parts of the saturated phase of our two-stream simulations. However, the saturated phase is generally dominated by nonlinear physics, thus linear theory should not be expected to give accurate predictions. Therefore, we restrict our comparison with \eqref{eq:A1A6LinThPred_gammato0} to the start at the saturated phase, before the created electron holes start to merge. Specifically, let us examine the time around when the peak average $\Ev$-field energy is reached, marked in \figref{fig:A1LinThComp_gammato0}a for the case where $n_b/n_e = 0.1$. Performing SR over the region where the instantaneous growth rate satisfies $\abs{\ga(t)}/\ope < 0.02$ near this peak for each two-stream simulation 
and comparing the SR value of $A_1$ with the value predicted by \eqref{eq:A1A6LinThPred_gammato0} yields \figref{fig:A1LinThComp_gammato0}b. As we can see, the agreement is good for weaker beam strengths, but becomes less accurate when approaching $n_b/n_e = 0.5$, where the physics involved are more nonlinear by virtue of the larger perturbation amplitudes.

If we instead consider the limit $\om_r \to 0$, corresponding to a non-oscillatory -- but possibly growing or decaying -- perturbation, we get
\begin{equation}
\begin{dcases}
    kA_6 = -\frac{1}{2} A_5 \qty( 1 + 3 \frac{\ope^2 + \ga^2}{k^2\vtb^2} ) \frac{k\vtb}{\ga} \\
    A_1 = -3 - \frac{\ope^2 + \ga^2}{k^2\vtb^2} + \qty[ -k A_2 + \frac{1}{2} k A_3 \qty( 1 + \frac{\ope^2 + \ga^2}{k^2\vtb^2} ) ] \frac{k\vtb}{\ga}.
\end{dcases}
\end{equation}
Similar to the case with $\ga \to 0$, the first equation allows for a solution where $A_6 = A_5 = 0$, consistent with the lack of wave propagation -- in fact, $A_4$ can also be set to zero. As for the second equation, if we insert inferred parameter values from the symmetric two-stream unstable setup, 
\begin{equation}
\begin{dcases}
    \ga = \SI{0.276}{\ope} \\
    \vtb = \SI{1.05e-2}{c} \\
    k \approx \kc = \SI{-67.4}{\de_e^{-1}},
\end{dcases}
\end{equation}
we get
\begin{equation}
    \frac{\ope^2 + \ga^2}{k^2\vtb^2} \approx 2.15 \text{ and } \frac{k\vtb}{\ga} \approx -2.56
\end{equation}
giving a prediction of
\begin{equation}
    A_1 \approx -5.15 + 2.56 \qty( k A_2 - 1.58 k A_3 ).
\end{equation}
The coefficient values found by SR in this case are
\begin{equation}
\begin{cases}
    A_1 = -3.90 \\
    A_2 = \SI{7.18e-3}{\de_e} \\
    A_3 = \SI{9.56e-3}{\de_e},
\end{cases}
\end{equation}
and if we insert our values for $A_2$ and $A_3$ into the approximate expression for $A_1$, we get $A_1 = -3.79$, which is reasonably accurate considering the quite broad $k$ spectrum.

\begin{figure}
    \centering
    \includegraphics[width=\textwidth]{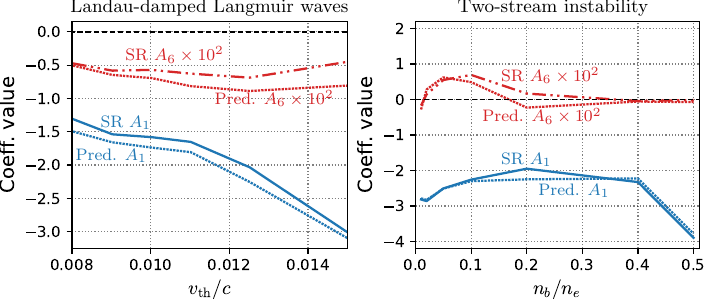}
    \caption{The values of $A_1$ and $A_6$ found by SR during the growth phase compared to the linear theory prediction given by \eqref{eq:A1A6LinThPred}, inserting the values of the other coefficients found by SR. The plot on the left contains the results from the Landau-damped Langmuir wave simulations, while the plot on the right contains the results from those with two-stream unstable setups.}
    \label{fig:A1A6LinThComp}
\end{figure}

Performing a similar comparison between the values of $A_1$ and $A_6$ found by SR during linear decay/growth, and the ones predicted by \eqref{eq:A1A6LinThPred} for all simulations, given the other coefficients as input, yields \figref{fig:A1A6LinThComp}. The agreement is decent for both the Landau damping simulations and the two-stream instability ones. Interestingly, the two-stream instability simulations agree even better with linear theory, despite their more broad-spectrum nature. This is likely due to a combination of several factors. The 
instantaneous decay rate in the Landau damping case oscillates throughout the decay, which means that assigning the $A_2$, $A_3$ and $A_6$ coefficients a single value for the entire decay phase is less accurate than doing the same for the growth phase in the two-stream simulations. 
In addition, the spacetime domain is larger in the two-stream simulations, yielding higher-resolution DFT spectra.

\subsection{A term $\propto n \vt V^2$}
\label{sec:results_brakingterm}
The terms making up the 6-term model are in some cases not the only ones found consistently by SR. In particular, a term $\propto n \vt V^2$ appears consistently in regressions over the growth phase when $\abs{\ga}$ is small -- at $\vtb/c < \SI{0.01}{}$ 
corresponding to $\abs{\ga}/\ope \ll \SI{0.15}{}$ for the Landau damping simulations and at $n_b/n_e < 0.4$, i.e.~$\abs{\ga} < \SI{0.27}{\ope}$, in the two-stream simulations. Like the terms in $\qo$, this term is $k$-odd, switching sign depending on the propagation direction of the waves. Notably, the cases where the term shows up are precisely those where one would expect a $k$-odd trapping-related term to show up, corresponding to high $\DtL/t_b$ and a clear wave propagation direction (i.e. high $\abs{\om/k}$).

\begin{figure}
    \centering
    \includegraphics[width=\textwidth]{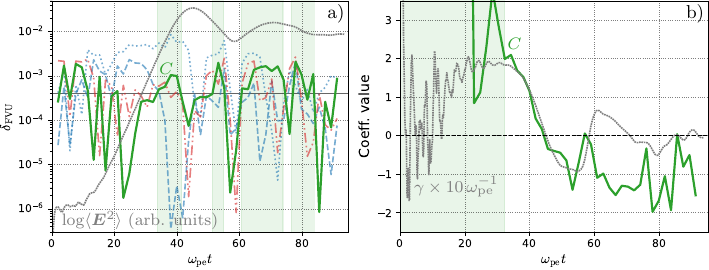}
    \caption{The variation over time of a) the importance of the braking term $C n \vt V^2$ as measured by $\dFVU$ and b) the coefficient value $C$ itself, for the two-stream unstable setup with $n_b/n_e = 0.1$. In panel a), we also show the $\dFVU$ values corresponding to the $A_2$, $A_3$ and $A_6$ terms for reference, illustrating how the braking term is of similar importance. We highlight the time ranges where the $\dFVU$ value is above the semi-arbitrary cutoff $\SI{4e-4}{}$, largely corresponding to decreasing $\abs{\ga(t)}$. We also show the time evolution of the $\Ev$-field energy in arbitrary units. In panel b), we show the instantaneous growth rate $\ga(t)$, illustrating its correlation with $C$. In the highlighted region, the prevalence of high-frequency noise and rapid oscillation in the growth rate causes the best-fit value of $C$ to oscillate wildly. To improve legibility, we thus only show $C$ at $\ope t > 20$.}
    \label{fig:7term_tEvol}
\end{figure}

Examining how the importance of this term evolves over time, as measured by $\dFVU$, we consistently find that it is almost as important as the $A_2$ and $A_6$ terms. At a few instances during the simulations, however, it is significantly more important than these two terms, and sometimes even more important than $A_3$. It is never as important as the $A_1$ and $A_5$ terms, however. Consistently, the periods where it is of higher importance occur towards the end of linear processes, when $\abs{\ga(t)}$ is decreasing, as can be seen in \figref{fig:7term_tEvol}a, which shows the case $n_b/n_e = 0.1$. Because of this, it might be reasonable to refer to this term as a kind of \say{braking term}, working to damp ongoing growth or decay.

As for the value of the coefficient itself (which we can call $C$), it correlates well with $\ga(t)$ starting in the latter half of the growth phase. This can be seen in \figref{fig:7term_tEvol}b, where the time evolution of the $C$ coefficient in the case where $n_b/n_e = 0.1$ is shown. Unlike terms $A_2$, $A_3$ and $A_6$, however, it only changes sign once for this value of $n_b/n_e$, going from positive in the growth phase to negative in the saturated phase. However, its magnitude continues to oscillate along with the growth-related terms even in the saturated phase. The high-frequency noise (and oscillation in $\ga$) in the beginning of the growth phase causes the best fit value to oscillate wildly during this time span, and because of this, we plot $C$ only after these oscillations start to die down. At other values of $n_b/n_e$, the time evolution of $C$ is similar, but the average value of the coefficient in the saturated phase, around which it oscillates, varies. More specifically, both the average value and the oscillation amplitude of $C$ seem to grow in absolute value as $n_b/n_e$ increases, until the term becomes unimportant at $n_b/n_e \to 0.5$ like the other $k$-odd terms. 

This behaviour can be partly explained by the fact that $n \vt V^2$ is second order in $r$ 
since $V \sim r \vp$ while both $n$ and $\vt$ have non-zero unperturbed values. More specifically, because of its second-order nature, we expect this term to matter only when the processes of interest deviate significantly from linearity.

\section{Discussion and conclusions}
\label{sec:conclusion}
Utilising methods from across the spectrum between physics fidelity on one hand and numerical tractability on the other is vital in exploring and understanding collisionless multi-scale plasma systems. In this context, collisionless fluid models play an important role, allowing global, long-timescale modelling of systems where this is not feasible with kinetic simulations. However, fluid models require closures to capture essential unresolved kinetic physics relevant to the plasma phenomena in question. Furthermore, theoretical closures are often derived by making idealised assumptions such as linearity or adiabatic invariance, which in many cases are broken by the dynamics of the system, motivating the use of data-driven approaches. 

We have performed a theoretical and numerical study of how sparse regression can be used to discover local heat flux closures in 1D electrostatic plasmas, examining Landau-damped Langmuir waves as well as two-stream instabilities. To ensure our inferred closures are able to capture kinetic effects, we generate our data using first-principles kinetic simulations -- specifically, the OSIRIS code. The closures identified by sparse regression regularly account for more than \SI{95}{\percent} of the variation in the heat flux, while remaining limited in complexity. Thus, we demonstrate the utility of an SR-based approach to systematic closure discovery.

As noted in \secref{sec:methods}, the high accuracy of the models found by SR suggests that our term library is large enough to capture the majority of the physics at play. To test whether there is room for improvement by exploring a larger function space (while having the closure remain local), one could employ neural network models and see how they perform compared to our SR models. No such analysis is included in this paper, however, since our models already reach FVU values of a few percent.
 
Besides a local approximation of the Hammett-Perkins closure, we consistently find several additional closure terms in both scenarios.
Notably, the three overall most important terms -- often accounting for over \SI{90}{\percent} of the variation in $q$ -- do not include the local Hammett-Perkins term. As one of these is a constant term, most of the variation in the heat flux divergence is captured by only two terms in the $q$ model: one $\propto n \vt^2 V$ and one $\propto n \vt^3$.
We further describe how the accuracy of the closure can be further improved by adding three more terms. These include a term $\propto n \vt^2 \pd_x \vt$, the local approximation of the Hammett-Perkins closure, along with terms $\propto \vt^3 \pd_x n$ and $\propto n \vt^2 \pd_x V$.
These terms are closely connected to the growth or decay of waves, their best-fit coefficients being approximately proportional to the growth rate.

Among these six terms, three are independent of propagation direction and three change sign depending on propagation direction (and are thus \say{$k$-even} and \say{$k$-odd}, respectively).
In a three-dimensional (3D) setting, this likely corresponds to the absence or presence of a unit vector in the wave propagation direction in the tensorial expression for the closure terms, so that terms with $k$-odd expressions would appear as e.g. $\propto\{ \hat{\boldsymbol{k}} \OmJ \}$ for some 2-tensor $\OmJ$.
This dependence on propagation direction for the $k$-odd terms also means that they can only exist when wave propagation or beam asymmetry breaks isotropy.

Having reached these results, we compared the closure coefficients with predictions from linear collisionless theory, overall with quite good agreement.
The two constraints imposed by linear theory give relationships between the coefficients of the various terms found by SR and the wave parameters $\om_r$, $\ga$ and $k$, as well as the plasma frequency $\ope$ and ambient thermal speed $\vtb$ of the plasma.
Not entirely unexpectedly, the appearance of frequency and wave number in these constraints suggests that fully capturing wave-plasma interactions requires a spatio-temporally non-local model.

Local approximations can still very much be utilised, however, providing one knows which parameter regimes are likely to be most relevant for the physics -- cf. the local approximation of the originally non-local Hammett-Perkins closure, which needs to be supplied with a characteristic wavenumber and heat diffusivity.
Already from the two constraints given by linear theory, various heuristics can be extracted depending on the parameter regime of interest.
Ultimately, all parameter dependencies should be made explicit and absorbed in the closure terms, so that the coefficients are parameter-independent, giving the closure as wide a range of applicability as possible. Further elucidating the parameter dependencies of the closure coefficients is left for future work.

After examining these six terms, which are all zeroth or first order in the perturbation amplitude $r = \tilde{n}/\bar{n}$, we also studied one additional term with non-negligible influence on $q$. 
This term,  $\propto n \vt V^2$, is second-order in $r$, and was found to mostly be important when growth or decay is slowing down.
The importance of each of these these seven terms was then investigated by examining their respective contributions to lowering the fraction of variance unexplained, or FVU, of the model.

Despite the high accuracy of the models described in this work, expanding the term library used by SR does merit some further investigation.
Apart from more complicated expressions involving $n$, $V$ and $\vt$, the $\Ev$ and $\Bv$ fields are also of interest -- especially for higher-dimensional non-electrostatic setups.
However, if the relevant physics is 2D or 3D, the number of relevant components of the vectors and tensors involved increases as well, necessitating even more careful selection of which terms to include in the SR term library.
There is also reason to explore alternative algorithms for sparse regression such as SINDy-PI \citep{Kaheman2020}, which generalises PDE-FIND to allow for implicit expressions for the quantity of interest $y$.
It should further be noted that using SR in Fourier space in order to directly identify non-local closures like the one originally proposed by Hammett and Perkins for Landau damping warrants more investigation.

As outlined in \secref{sec:introduction}, exploring ways of systematically discovering fluid closures is chiefly motivated by the enormous computational complexity of accurately modelling multi-scale processes in plasmas.
In particular, one of the main envisioned use cases for closures of the type presented in this paper is sub-grid scale modelling of rapid, small-scale processes within larger simulations.
When modelling instabilities in this way, the far future limit of the saturated regime (relative to the time scales of the instability) is the most important regime to model correctly.

Finally, evaluating the performance of -- and fully benefiting from  -- data-driven closures of this type requires a flexible implementation of closure terms in collisionless fluid solvers, such as the 10-moment solver of Gkeyll \citep{Hakim2006,Hakim2008}, and comparing the results with equivalent kinetic simulations.
Flexible closure prescription would also be a requirement for e.g. the robust -- and challenging -- approach demonstrated by \citet{Joglekar_2023}. In that work, the free parameters of a fluid closure are learned by neural network models, and a differentiable fluid solver is used,  enabling the calculation of the loss function gradient with respect to the neural network weights. The loss function in this case quantifies the difference between the long time predictions of physics observables as calculated by a kinetic and the fluid solver. The complexity of this training process limits the number of free parameters. We envision that the SR approach we explored here can inform a good (and interpretable) starting point for closure parametrisation. The parametric dependences of the coefficients of the model terms may then be be fine-tuned with a differentiable fluid solver.   

It should be noted that implementing fluid closures in existing fluid codes is far from trivial, due to the possibility of unphysical instabilities arising unless the closure is chosen with this aspect in mind; indeed arbitrarily small errors in model coefficients may lead to numerical instability. Enforcing long-time boundedness of discovered models is possible in systems with quadratic nonlinearities \cite{KaptanogluStability}, but it is difficult more generally.   
In fact, many closures (e.g. the various existing ad-hoc relaxation closures \citep{Wang2015,Ng2015}) are in part used because of their ability to \say{diffuse} anisotropies, reducing the complexity of dynamics and increasing the stability of the simulation. In our context, the signs of the closure coefficients (in particular that of the Hammett-Perkins-like term $A_3$) found in the Landau damping scenario are such as to increase entropy and thus provide stability. However, they correspond to an instability in the two-stream unstable scenario. 
It should be emphasised that having such fundamentally instability-driving terms in our closure in this case is necessary to accurately model the growth phase solely because we are modelling this physically unstable scenario using only a single electron species. Treating the counter-streaming populations as separate fluid species may be applied to resolve this shortcoming, which is the subject of ongoing investigations.    

Notably, however, closures found by SR such as the ones described in this paper hold an advantage when it comes to ensuring stability as compared to those based on e.g.~neural networks.
This is due to their interpretability and low complexity, which makes it possible to study the properties of the closures analytically -- something which is generally not feasible for neural networks.
    
\section*{Acknowledgements} 
The authors are grateful to K.~Steinvall, D.~Graham and T.~F\"{u}l\"{o}p for fruitful discussions.
The computations used the OSIRIS particle-in-cell simulation code, and were enabled by resources provided by the National Academic Infrastructure for Supercomputing in Sweden (NAISS), partially funded by the Swedish Research Council through grant agreement No.~2022-06725.

\section*{Funding} 
The work was supported by the Knut and Alice Wallenberg foundation (Dnr.~2022.0087) and the Swedish Research Council (Dnr.~2021-03943).  
This work was also supported by the National Science Foundation Grants No.~PHY-2018087 and PHY-2018089. 

\section*{Declaration of Interests}
Competing interests: The authors declare none.

\appendix
    


\section{Constraints from linear collisionless theory}
\label{app:LinearCollisionlessTheory}
In this section we construct constraint relations between terms appearing in a local expression for the heat flux, which then can be applied to the closure terms found using SR. 

Our starting point is the Vlasov-Maxwell system, i.e.
\begin{equation}
\begin{dcases}
    \pd_t f_\sg + \dirdvJ{\vv} f_\sg + \dirdvJ{\frac{q_\sg}{m_\sg}\qty(\Ev + \vv \times \Bv)}_{\!\vv} f_\sg = 0\\
    \divJ{\Ev} = \frac{1}{\ve_0} \sum_\sg q_\sg n_\sg \\
    \divJ{\Bv} = 0 \\
    \curlJ{\Ev} = - \pd_t \Bv \\
    \curlJ{\Bv} = c^{-2} \pd_t \Ev + \mu_0 \sum_\sg q_\sg n_\sg \Vv_{\!\!\sg},
\end{dcases}
\end{equation}
which is the most accurate self-consistent continuum description of collisionless plasmas. However, since this coupled system of PDEs is very expensive to solve over large domains while retaining high resolution, it is often necessary to simplify it. Most relevant for us is the fact that one can integrate the Vlasov equation over velocity space to instead get the fluid equations \cite{Grad49,Levermore96}. Truncating these after the pressure equation yields the system sometimes referred to as the 10-moment model \citep{Wang2015}
\begin{equation}
\begin{dcases}
    \pd_t n_\sg + \divJ{\qty(n_\sg \Vv_{\!\!\sg})} = 0 \\
    n_\sg \qty( \pd_t + \dirdvJ{\Vv_{\!\!\sg}} ) \Vv_{\!\!\sg} + \divJ{\pJ_\sg} = \frac{q_\sg}{m_\sg} n_\sg \qty(\Ev + \Vv_{\!\!\sg} \times \Bv) \\
    \pd_t \pJ_\sg + \divJ{ \qty(\Vv_{\!\!\sg} \pJ_\sg) } + 2 \qty{ \dirdvJ{\pJ_\sg} \Vv_{\!\!\sg} } + \divJ{\qJ_\sg} = \frac{2q_\sg}{m_\sg} \qty{ \pJ_\sg \times \Bv }, 
    \label{fluideqs}
\end{dcases}
\end{equation}
which needs to be closed by supplying an additional expression for $\qJ_\sg$ (or $\divJ{\qJ_\sg}$) in terms of the lower moments. Here, $\qty{\cdot}$ denotes symmetrisation, so that e.g.
\begin{equation}
    \qty{\av \bv^{(2)}} = \frac{1}{3} \qty( \av \bv^{(2)} + \bv \av \bv + \bv^{(2)} \av ),
\end{equation}
and $\pJ_\sg \times \Bv$ should be interpreted as the 2-tensor with elements $\qty[\pJ_\sg \times \Bv]_{ij} = \lc_{jkl} \, p_{\sg ik} B_l$.

In 1D electron-proton setups like those of interest to us, where only the electron dynamics are important, the 10-moment fluid model (sans closure) together with Maxwell's equations simplifies to
\begin{equation}
\begin{dcases}
    n \qty( \pd_t + V \pd_x ) V + \pd_x  p = -\frac{e}{m_e} n E \\
    \qty(\pd_t + V \pd_x) p + 3 p \pd_x V + \pd_x q = 0 \\
    \pd_x E = \frac{e}{\ve_0} \qty( \bar{n} - n) \\
    \pd_t E = \frac{e}{\ve_0} n V.
\end{dcases}
\label{eq:1D_fluidMxw}
\end{equation}
Note that the two remaining Maxwell's equations imply the continuity equation. Here, we have taken the ions to be immobile with number density equal to the average electron density $\bar{n}$ to ensure quasi-neutrality. Now, let us consider a small wave-like perturbation around equilibrium in the CoM frame, i.e.
\begin{equation}
\begin{cases}
    n = \bar{n} + \tilde{n} e^{i(kx - \om t)} \\
    V = \tilde{V} e^{i(kx - \om t)} \\
    p = n v_\text{th}^2, \; \vt = \vtb + \vtt e^{i(kx - \om t)} \\
    q = \bar{q} + \tilde{q} e^{i(kx - \om t)} \\
    E = \tilde{E} e^{i(kx - \om t)}.
    \label{ansatz1}
\end{cases}
\end{equation}
Here, we assume the wavenumber $k$ to be real, but the frequency $\om = \om_r + i\ga$ is allowed to be complex, $\ga$ being the growth rate. Of course, one could equally well set $\om = \om_r - i\ga$, taking $\ga$ as the decay rate.

Inserting ansatz (\ref{ansatz1}) into \eqref{eq:1D_fluidMxw} and keeping only terms up to first order in the perturbations, we get the relations making up 1D linear collisionless theory:
\begin{equation}
\begin{dcases}
    -i\om \bar{n} \tilde{V} + ik \qty( \tilde{n} \vtb^2 + 2\bar{n} \vtb \vtt ) = - \frac{e}{m_e} \bar{n} \tilde{E} \\
    -i\om \qty( \tilde{n} \vtb^2 + 2\bar{n} \vtb \vtt ) + 3ik \bar{n} \vtb^2 \tilde{V} + ik \tilde{q} = 0 \\
    ik \tilde{E} = - \frac{e}{\ve_0} \tilde{n} \\
    -i\om \tilde{E} = \frac{e}{\ve_0} \bar{n} \tilde{V},
\end{dcases}
\end{equation}
or equivalently
\begin{equation}
\begin{dcases}
    \tilde{E} = i \frac{e}{k \ve_0} r \bar{n} \\
    \tilde{V} = r \vp \\
    \vtt = -\frac{1}{2} \qty( 1 + \frac{\ope^2 - \om^2}{k^2 \vtb^2} ) r \vtb \\
    \tilde{q} = - \qty( 3 + \frac{\ope^2 - \om^2}{k^2 \vtb^2} ) r \bar{n} \vtb^2 \vp.
\end{dcases}
\label{eq:LinearTheoryCriteria}
\end{equation}
To make the notation neater, we have introduced the (complex) phase velocity $\vp = \om / k$ and the shorthand notation $r$ for $\tilde{n} / \bar{n}$, quantifying the amplitude of the perturbation. We have also introduced the electron plasma frequency $\ope = \sqrt{e^2 \bar{n} / m_e \ve_0}$. Note that when the ion dynamics is negligible, any heat flux closure would need to agree with the expression for $\tilde{q}$ to first order in $r$ in order to be viable for modelling weak wave-like perturbations.
    


\subsection{Evaluating the 6-term closure}
For a closure to be consistent with theory, the heat flux perturbation amplitude $\tilde{q}$ given by the closure must agree with the final part of \eqref{eq:LinearTheoryCriteria}, i.e.
\begin{equation}
    \tilde{q} = - \qty( 3 + \frac{\ope^2 - \om^2}{k^2 \vtb^2} ) r \bar{n} \vtb^2 \vp,
\label{eq:LinearTheory_qCriterion}
\end{equation}
to first order in $r$. To see whether this is indeed the case, let us calculate the contribution to $\tilde{q}$ from each of the terms in the 6-term model. For the terms in $\qe$, we get
\begin{equation}
\begin{dcases}
    A_1 \qty( \tilde{n} \vt^2 V + 2 n \vt \vtt V + n \vt^2 \tilde{V} ) &= A_1 r \bar{n} \vtb^2 \vp \\
    ik A_2 \qty( 3 \tilde{n} \vt^2 \vtt + \tilde{n} \vt^3 ) &= ik A_2 r \bar{n} \vtb^3 \\
    ik A_3 \qty( \tilde{n} \vt^2 \vtt + 2 n \vt \vtt^2 + n \vt^2 \vtt ) &= -\frac{1}{2}ik A_3 \qty( 1 + \frac{\ope^2 - \om^2}{k^2\vtb^2} ) r\bar{n} \vtb^3,
\end{dcases}
\end{equation}
keeping only terms up to first order in $r$. In other words,
\begin{equation}
    \tilde{q}_\text{even} = -\qty[ -A_1 + ik \qty( \frac{1}{2} A_3 - A_2) \frac{\vtb}{\vp} + \frac{1}{2} ik A_3 \frac{\vtb}{\vp} \frac{\ope^2 - \om^2}{k^2\vtb^2} ] r \bar{n} \vtb^2 \vp.
\end{equation}
As for the terms in $\qo$, the constant term $A_4$ does not contribute to $\tilde{q}$, but the contribution from the other two terms is non-zero:
\begin{equation}
\begin{dcases}
    A_5 \qty( \tilde{n} \vt^3 + 3 n \vt^2 \vtt ) &= - \frac{1}{2} A_5 \qty( 1 + 3 \frac{\ope^2 - \om^2}{k^2\vtb^2} ) r \bar{n} \vtb^3 \\
    ik A_6 \qty( \tilde{n} \vt^2 \tilde{V} + 2 n \vt \vtt \tilde{V} + n \vt^2 \tilde{V} ) &= ik A_6 r \bar{n} \vtb^2 \vp. \\
\end{dcases}
\end{equation}
Thus, the contribution from $\qo$ is
\begin{equation}
    \tilde{q}_\text{odd} = - \qty[ \frac{1}{2} A_5 \frac{\vtb}{\vp} - ik A_6 + \frac{3}{2} A_5 \frac{\vtb}{\vp} \frac{\ope^2 - \om^2}{k^2\vtb^2} ] r \bar{n} \vtb^2 \vp.
\end{equation}
Defining
\begin{equation}
    \Phi(\om,k) = \frac{\ope^2 - \om^2}{k^2\vtb^2}
\end{equation}
as well as
\begin{equation}
    \al = \Re \Phi = \frac{\ope^2 + \ga^2 - \om_r^2}{k^2\vtb^2}, \quad \be = -\Im \Phi = \frac{2 \om_r \ga}{k^2\vtb^2},
\end{equation}
demanding \eqref{eq:LinearTheory_qCriterion} hold for our closure is equivalent to demanding
\begin{equation}
    \qty(3 + \Phi) \om = - (A_1 + ik A_6) \om + \qty[ \frac{1}{2} A_5 + ik \qty( \frac{1}{2} A_3 - A_2 ) ] k\vtb + \frac{1}{2} \qty( 3 A_5 + ik A_3 ) k\vtb \Phi,
\end{equation}
or equivalently
\begin{equation}
\begin{dcases}
    \qty(3 + \al) \om_r + \be\ga = -A_1 \om_r + k A_6 \ga + \frac{1}{2} \qty[ k A_3 \be + A_5 (1 + 3\al) ] k\vtb \\
    -\qty(3 + \al) \ga + \be \om_r = A_1 \ga + k A_6 \om_r + \qty[ kA_2 - \frac{1}{2} kA_3 \qty(1 + \al) + \frac{3}{2} A_5 \be ] k\vtb.
\end{dcases}
\end{equation}
If we now define
\begin{equation}
    \Phi_\pm = \frac{\ope^2 \pm \abs{\om}^2}{k^2\vtb^2}
\end{equation}
and solve for $A_1$ and $kA_6$; the result can be simplified to the form of \eqref{eq:A1A6LinThPred},
giving us two constraints on the coefficients which we can check.
    


\section{Demonstration: recovery of the momentum equation}
\label{app:SR_Efieldalignment}
While the main use of sparse regression in this paper is to discover unknown approximate relations between the heat flux and lower order fluid quantities, it is useful to verify that our workflow is able to identify known exact relations that the simulation data must obey. To illustrate that this is indeed the case, we here show how one can recover the density-normalised version of the 1D momentum equation for electrons, 
\begin{equation}
    \pd_t V = - V \pd_x V - T \pd_x \ln \frac{n}{\bar{n}} - \pd_x T  -\frac{e}{m_e} E,
    \label{eq:1dmomeq_demo}
\end{equation}
from our two-stream simulation data. Note that the arbitrary normalisation of $n$ to the unperturbed total electron density $\bar{n}$ does not affect the the logarithmic derivative. In the sparse regression we now set $\pd_t V$ as our target variable $y$ and use a term library including all products of up to two terms from the set consisting of $n$, $V$, $T$ and $E$, their spatial derivatives and a constant term.

\begin{figure}
    \centering
    \includegraphics[width=\textwidth]{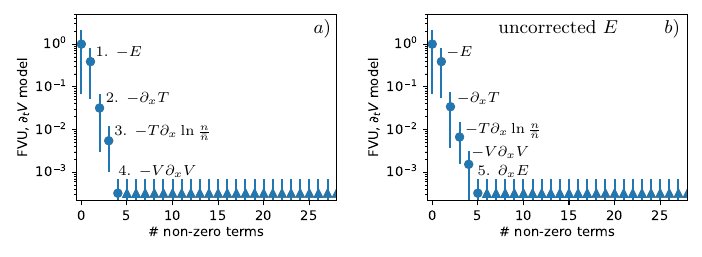}
    \caption{The sequence of models found by sparse regression to approximate $\partial_t V$, recovering the momentum equation as given in \eqref{eq:1dmomeq_demo}. The consistently found models are labelled by the new consistently found term which is added to the model compared to the next most simple one. The coefficients for all terms in each model are provided in \tabref{tab:recovery} (with the same numbering of models as here). Circles, unlike triangles, denote consistently found terms, and blue (orange) marker color corresponds to testing (training) data (note that they overlap almost completely here). The two subfigures show the results a) when using $\Ev$-field data which has been corrected for the half-cell grid shift with respect to the fluid quantities, which finds only the expected terms, and b) when using uncorrected $\Ev$-field data, which also finds a non-physical $\partial_x E$ term adding the correction back in.}
    \label{fig:SR_momeqdemo}
\end{figure}

We work with the $n_b/n_e = 0.1$ two-stream instability simulation, sampling $\sim \SI{2.5}{\percent}$ of the data in each of the 10 cross-validation folds. This time we take samples from almost the entire time domain, and not just e.g. the growth phase. The sparse regression algorithm yields the sequence of models for $\pd_t V$ shown in figure \figref{fig:SR_momeqdemo}a.

As shown by the marker shapes -- indicating whether models contain consistently the same terms (circles) or different ones (triangles) -- SR correctly recovers the momentum equation and finds no further terms consistently. That the model is complete at four terms is also supported by the fact that the accuracy plateaus after this point, at an FVU of $\sim \SI{3.26e-4}{}$. This signifies that $\sim \SI{99.97}{\percent}$ of the variation in $\pd_t V$ can be explained by the 1D momentum equation as given in \eqref{eq:1dmomeq_demo}. SR also gives us the coefficients of the terms in the equation with an error of less than half a percent; these are listed in \tabref{tab:recovery}.

\begin{table}
    \centering
    \begin{tabular}{c|cccc|c}
        \toprule
        Model & $-E$ & $-\partial_x T$ & $-T\partial_x \ln \frac{n}{\bar{n}}$ & $-V\partial_x V$ & $\partial_x E$  \\
        \midrule
        1. & $0.739$ & ${}$ & ${}$ & ${}$ & ${}$  \\ 
        2.  & $1.234$ & $1.138$ & ${}$ & ${}$ & ${}$  \\ 
        3.  & $0.985$ & $0.977$ & $1.006$ & ${}$ & ${}$  \\
        4.  & $0.999$ & $0.998$ & $0.996$ & $0.997$ & ${}$  \\
        \midrule
        5. (uncorrected $E$)  & $1.000$ & $0.998$ & $0.991$ & $0.991$ & $5.00\times 10^{-4}$  \\
        \bottomrule
    \end{tabular}
    \caption{\label{tab:recovery} Optimal coefficient values for the terms in the various models found by sparse regression to approximate the momentum equation as given in \eqref{eq:1dmomeq_demo}. The numbering of the models is the same as in \figref{fig:SR_momeqdemo}. The theoretical value of almost all coefficients multiplying the terms shown in the first row is $1$. The only exception is the non-physical $\pd_x E$ term, which is only found consistently when the staggering of the $E$ field data is not corrected for (model 5.). This coefficient should have the value $5\times 10^{-4}$ to provide a first order correction for the half cell size shift of the field data.}
\end{table}

Next, we illustrate the importance of appropriately aligning the $\Ev$-field data with the fluid data in space. Misalignment between particle and field data naturally occurs in PIC schemes using a staggered Yee grid \citep{Yee1966}, where the $\Ev$- and $\Bv$-field nodes appear in cell edges and cell faces, respectively, while the particle data is often cell-centred.\footnote{In addition, the leap-frog type time integration scheme results in momentum and position information available half time step apart which, unless corrected for (as in the version of OSIRIS we use), may also lead to a reduced accuracy. Note that, unlike the spatial misalignment, the temporal one cannot be corrected by post-processing; the correction needs to be made before the fluid quantities are computed by the simulation code.} This discrepancy then propagates through to any fluid and electromagnetic field data exported from the simulation. 

In \figref{fig:SR_momeqdemo}b we show the result of performing SR on the same simulation data without correcting for the misalignment of the electric field data. The first four consistently found terms are the same as before, but now this 4-term model is less accurate, with an FVU $>10^{-3}$. A fifth term $\propto \pd_x E$ is also consistently identified, however, and if retained increases the accuracy to match that achieved with corrected $\Ev$-field data. This term corresponds to performing a first order Taylor expansion to evaluate $E$ half a grid cell further towards $-x$, since $\eval{E}_{x - \frac{1}{2} \De x} = \eval{E}_x - \frac{1}{2} \De x \eval{\pd_x E}_x$. Specifically, if we insert our spatial resolution $\De x = \SI{e-3}{\de_e}$, we find that
\begin{equation}
    E_\text{aligned} = E_\text{shifted} - \SI{5e-4}{\de_e} \pd_x E_\text{shifted},
\end{equation}
corresponding exactly to the correction introduced by SR, since $\de_e$ is our unit of length.

\bibliographystyle{jpp}
\bibliography{ref}

\end{document}